\documentclass[twoside,twocolumn,9pt]{article}
\usepackage{colortbl}
\usepackage{extsizes}
\usepackage{amsmath,amssymb}
\usepackage[super,sort&compress,comma]{natbib} 
\usepackage[version=3]{mhchem}
\usepackage[left=1.5cm, right=1.5cm, top=1.785cm, bottom=2.0cm]{geometry}
\usepackage{balance}
\usepackage{times,mathptmx}
\usepackage{sectsty}
\usepackage{graphicx} 
\usepackage{lastpage}
\usepackage{adjustbox}
\usepackage[format=plain,justification=justified,singlelinecheck=false,font={stretch=1.125,small,sf},labelfont=bf,labelsep=space]{caption}
\usepackage{float}
\usepackage{fancyhdr}
\usepackage{fnpos}
\usepackage[english]{babel}
\addto{\captionsenglish}{%
  
}
\usepackage{array}
\usepackage{droidsans}
\usepackage{charter}
\usepackage[T1]{fontenc}
\usepackage[usenames,dvipsnames]{xcolor}
\usepackage{setspace}
\usepackage[compact]{titlesec}
\usepackage{hyperref}
\DeclareUnicodeCharacter{0308}{~}

\usepackage{gensymb} 
\definecolor{cream}{RGB}{222,217,201}
\begin{document}

\pagestyle{fancy}
\thispagestyle{plain}
\fancypagestyle{plain}{

\renewcommand{\headrulewidth}{0pt}
}

\makeFNbottom
\makeatletter
\renewcommand\LARGE{\@setfontsize\LARGE{15pt}{17}}
\renewcommand\Large{\@setfontsize\Large{12pt}{14}}
\renewcommand\large{\@setfontsize\large{10pt}{12}}
\renewcommand\footnotesize{\@setfontsize\footnotesize{7pt}{10}}
\makeatother

\renewcommand{\thefootnote}{\fnsymbol{footnote}}
\renewcommand\footnoterule{\vspace*{1pt}%
\color{cream}\hrule width 3.5in height 0.4pt \color{black}\vspace*{5pt}} 
\setcounter{secnumdepth}{5}

\makeatletter 
\renewcommand\@biblabel[1]{#1}            
\renewcommand\@makefntext[1]%
{\noindent\makebox[0pt][r]{\@thefnmark\,}#1}
\makeatother 
\renewcommand{\figurename}{\small{Fig.}~}
\sectionfont{\sffamily\Large}
\subsectionfont{\normalsize}
\subsubsectionfont{\bf}
\setstretch{1.125} 
\setlength{\skip\footins}{0.8cm}
\setlength{\footnotesep}{0.25cm}
\setlength{\jot}{10pt}
\titlespacing*{\section}{0pt}{4pt}{4pt}
\titlespacing*{\subsection}{0pt}{15pt}{1pt}

\fancyfoot{}
\fancyfoot[LO,RE]{\vspace{-7.1pt}\includegraphics[height=9pt]{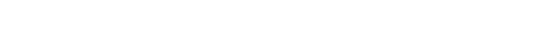}}
\fancyfoot[CO]{\vspace{-7.1pt}\hspace{13.2cm}\includegraphics{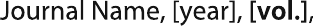}}
\fancyfoot[CE]{\vspace{-7.2pt}\hspace{-14.2cm}\includegraphics{head_foot/RF}}
\fancyfoot[RO]{\footnotesize{\sffamily{1--\pageref{LastPage} ~\textbar  \hspace{2pt}\thepage}}}
\fancyfoot[LE]{\footnotesize{\sffamily{\thepage~\textbar\hspace{3.45cm} 1--\pageref{LastPage}}}}
\fancyhead{}
\renewcommand{\headrulewidth}{0pt} 
\renewcommand{\footrulewidth}{0pt}
\setlength{\arrayrulewidth}{1pt}
\setlength{\columnsep}{6.5mm}
\setlength\bibsep{1pt}

\makeatletter 
\newlength{\figrulesep} 
\setlength{\figrulesep}{0.5\textfloatsep} 

\newcommand{\topfigrule}{\vspace*{-1pt}%
\noindent{\color{cream}\rule[-\figrulesep]{\columnwidth}{1.5pt}} }

\newcommand{\botfigrule}{\vspace*{-2pt}%
\noindent{\color{cream}\rule[\figrulesep]{\columnwidth}{1.5pt}} }

\newcommand{\dblfigrule}{\vspace*{-1pt}%
\noindent{\color{cream}\rule[-\figrulesep]{\textwidth}{1.5pt}} }

\makeatother

\twocolumn[
  \begin{@twocolumnfalse}
  	{\includegraphics[height=30pt]{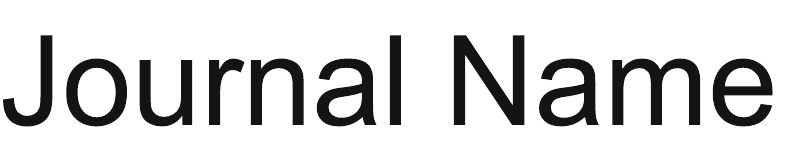}\hfill%
  		\raisebox{0pt}[0pt][0pt]{\includegraphics[height=55pt]{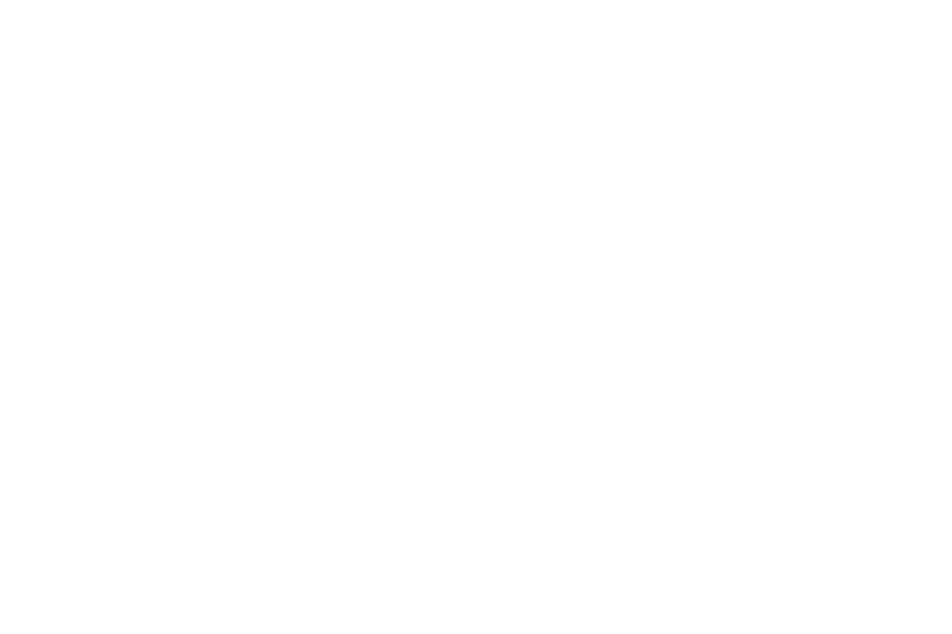}}%
  		\\[1ex]%
  		\includegraphics[width=18.5cm]{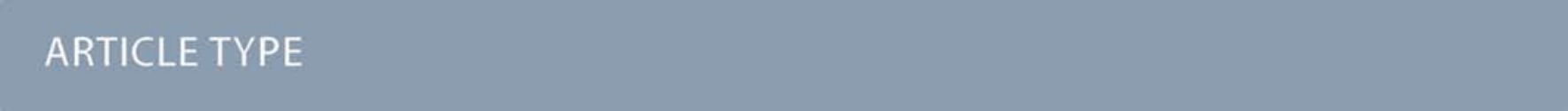}}\par
\vspace{1cm}
\sffamily
\begin{tabular}{m{4.5cm} p{13.5cm} }

\includegraphics{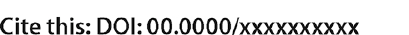} & \noindent\LARGE{\textbf{Theoretical evaluation of oxynitride, oxyfluoride and nitrofluoride perovskites with promising photon absorption properties for solar water splitting$^\dag$}} \\
\vspace{0.3cm} & \vspace{0.3cm} \\

 & \noindent\large{Manjari Jain\textit{$^{a}$}, Deepika Gill, Sanchi Monga and Saswata Bhattacharya$^{\ast}$\textit{$^{a}$}} \\

\includegraphics{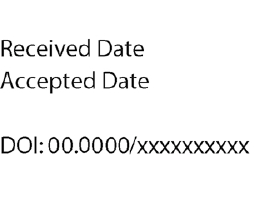} & \noindent\normalsize{Photocatalytic water splitting represents a very promising but at the same time very challenging contribution to a clean and renewable route to produce hydrogen fuel. Developing efficient and cost-effective photocatalysts for water splitting is a growing need. For this purpose, semiconductor photocatalysts have attracted much more attention due to their stability and low manufacturing cost. Here, we have systematically applied several state-of-the-art advanced first-principles-based methodologies, viz., hybrid density functional theory, many-body perturbation theory (G$_0$W$_0$) and density functional perturbation theory (DFPT), to understand the electronic structure properties of ABX$_2$Y perovskites. We have chosen the vast composition space of ABX$_2$Y type perovskites where A and B are cations and X and Y can be nitrogen, oxygen, or fluorine. These perovskites exhibit direct band gaps ranging from 1.6 to 3.3 eV. Further, to evaluate the feasibility of the visible light catalytic performance, we calculate the structural, electronic, and optical properties of ABX$_2$Y perovskites. In addition, 
from hydrogen evolution reaction (HER) and oxygen evolution reaction (OER) mechanism, BaInO$_2$F, InSnO$_2$N, CsPbO$_2$F and LaNbN$_2$O are found as probable photocatalysts.} \\
\end{tabular}

 \end{@twocolumnfalse} \vspace{0.6cm}

]

\renewcommand*\rmdefault{bch}\normalfont\upshape
\rmfamily
\section*{}
\vspace{-1cm}

\footnotetext{\textit{$^{a}$Department of Physics, Indian Institute of Technology Delhi, New Delhi, India. Fax: 91 11 2658 2037; Tel: 91 11 2659 1359; E-mail: saswata@physics.iitd.ac.in}}




\section{Introduction}
The photocatalytic water splitting has become an intense research area and an excellent way to capture and store energy from the sun~\cite{PMID:27994241,C4EE03869J,PMID:29123062,D1NA00154J}. Over recent decades, the number of applications based on photocatalysis increased acutely~\cite{D0TA10781F,C2EE23482C,C8TC05749D}. Although a wide range of materials has been developed for photocatalytic performance under visible light, most can only absorb light at a wavelength of less than 500 nm, so only a small portion of the spectrum can be utilized~\cite{B800489G,C3CS60378D,C3CC49279F,C5TA10612E}. Therefore, the design of a semiconductor for the efficient generation of solar fuel requires a suitable band gap (1.5 $\leq$ \textit{E}$_\textrm{g}$ $\leq$ 2.8 eV) to efficiently absorb visible light, high dielectric constant, high charge carrier mobility and suitable band positions in order to perform the hydrogen and oxygen evolution half-reactions (HER/OER)~\cite{doi:10.1021/acscatal.7b02662,C5TA06983A}.\\
So far, most of the existing photocatalysts are oxides~\cite{doi:10.1021/cm7024203}. However, they have too large band gap to absorb visible light~\cite{doi:10.1021/cm7024203,doi:10.1021/jp070911w}. This is mainly due to a very low valence band (VB) energy which comes from the 2p orbitals of the oxygen atoms~\cite{doi:10.1021/jp070911w}. A requirement for visible light induced photocatalysts is that the optimum band gap energy should be less than 3 eV~\cite{B417052K}. In order to solve this problem, non-oxides such as nitrides and sulfides have been proposed as their VB position is usually higher in energy~\cite{C9TA03116B,doi:10.1021/acscatal.1c03737}. Nitrogen is less electronegative than oxygen that leads to reduction in the band gap. Hence, the optical gap overlaps with the solar spectrum which makes this class of materials interesting for application as solar absorbers and visible light-driven photocatalysts~\cite{doi:10.1063/1.5140056}. Many useful oxynitrides have been reported, such as: (i) CaTaO$_2$N and LaTaON$_2$ are non toxic solid solutions, (ii) BaTaO$_2$N has a high dielectric constant and acts as a photocatalyst in water decomposition, and (iii) EuNbO$_2$N is ferromagnetic and show colossal magneto resistance~\cite{Yang2011,D0TA02136A,C5MH00046G}. Also, it has been reported that LaTaON$_2$, LaTiO$_2$N, SrTaO$_2$N and BaTaO$_2$N drive the HER and OER half-reactions and can be used in Z-scheme configurations capturing photons in the 600 to 750 nm range~\cite{Nurlaela2016,ABE2010179,C5EE01434D,D0TA08117E,D1TA05052D}. Recently, Zhang \textit{et al.} reported that CoO$_x$ modified LaTiO$_2$N has a high quantum efficiency of 27\% at 440 nm towards water oxidation~\cite{doi:10.1021/ja301726c}. LaMg$_x$Ta$_{1-x}$O$_{1+3x}$N$_{2-3x}$ is found to be the first oxynitride to utilize 600 nm photons in steady-state overall water splitting~\cite{C5EE01434D}. Moreover, various oxyfluoride perovskites such as BaFeO$_2$F, SrFeO$_2$F and PbFeO$_2$F have been discovered. These perovskites exhibit magnetic ordering until a temperature of around 645 K, 685 K, and >500 K, respectively~\cite{HEAP2007467,doi:10.1021/cm048125g,SrFeO2F}. Multiferroic behavior was also shown by iron-based oxyfluoride perovskites~\cite{doi:10.1021/acs.inorgchem.8b01253}. Also, Samir \textit{et al.} proposed the potential existence of a nitrofluoride (LaZrN$_2$F) from the first-principles based DFT approach. They showed that LaZrN$_2$F composition exhibits semiconducting properties with iono-covalent behavior~\cite{MATAR201077}. However, despite several research endeavors in the field of oxynitride, oxyfluoride and nitrofluoride perovskites, no such in-depth theoretical work is available to address their electronic, optical, dielectric properties and their application in photocatalytic water splitting.\\
In this work, we have used various advanced state-of-the-art first-principles based methodologies under the framework of density functional theory (DFT)~\cite{kohn1965self,hohenberg1964inhomogeneous}, many body perturbation theory (MBPT)~\cite{jiang2012electronic, fuchs2008efficient} and density functional perturbation theory (DFPT)~\cite{PhysRevB.73.045112,choudhary2020high} to provide a comprehensive computational study of oxynitride, oxyfluoride, and nitrofluoride perovskites. Our objectives are: (i) to provide a list of compositions that could be likely experimentally synthesized in the perovskite phase, (ii) to study their electronic structure, (iii) to study their optical and excitonic properties, and (iv) to find their application in photocatalytic water splitting. To achieve this goal, we report calculations for the theoretical electrocatalytic HER and OER overpotentials using DFT.
 
\section{Methodology}
We have performed a systematic study to explore the structural, electronic and optical properties using DFT and beyond approaches under the framework of MBPT.  All calculations are performed with Projector Augmented Wave (PAW) potentials as implemented in Vienna $\textit{ab initio}$ simulation package (VASP)~\cite{kresse1996efficiency,kresse1999ultrasoft}. All the structures are optimized using generalized gradient approximation (GGA) as implemented in PBE~\cite{perdew1996generalized} exchange-correlation ($\epsilon_{\textrm{xc}}$) functional until the forces are smaller than 0.001 eV/\AA. The $\Gamma$-centered 2$\times$2$\times$2 k-mesh sampling is employed for optimization calculations (optimized structures are shown in Fig.~\ref{1}). The electronic self-consistency loop convergence is set to 0.01 meV, and the kinetic energy cutoff is set to 600 eV for plane wave basis set expansion. To explore the electronic properties, hybrid $\epsilon_{\textrm{xc}}$ functional (HSE06)~\cite{krukau2006influence,heyd2003hybrid} is used. For all the energy calculations 4$\times$4$\times$4 k-mesh has been used. The phonon calculations are performed with 2$\times$2$\times$2 supercells using the PHONOPY package~\cite{TOGO20151,PhysRevB.78.134106}. Note that the surfaces (100 plane) were constructed by cleaving the fully optimized bulk structure with the lowest energy. Further, surface geometries are optimized with a force convergence threshold of 0.001 eV/\AA. Reciprocal space is sampled by 4$\times$4$\times$1 k-mesh for the (100) surface.  A 20 \AA\: vacuum is added to avoid artificial interaction between periodic images, and a dipole correction is included. The two-body Tkatchenko-Scheffler vdW scheme has been used to account for van der Waals interactions~\cite{PhysRevLett.108.236402,PhysRevLett.102.073005}. Note that the spin-orbit coupling (SOC) is not taken into consideration because it has negligible affect on the electronic structure of ABX$_2$Y perovskites (see section I of supplemental information (SI)).
\section{Results and Discussion}
\subsection{Crystal Structure}
\begin{figure}[h] 
	\centering
	\includegraphics[width=0.45\textwidth]{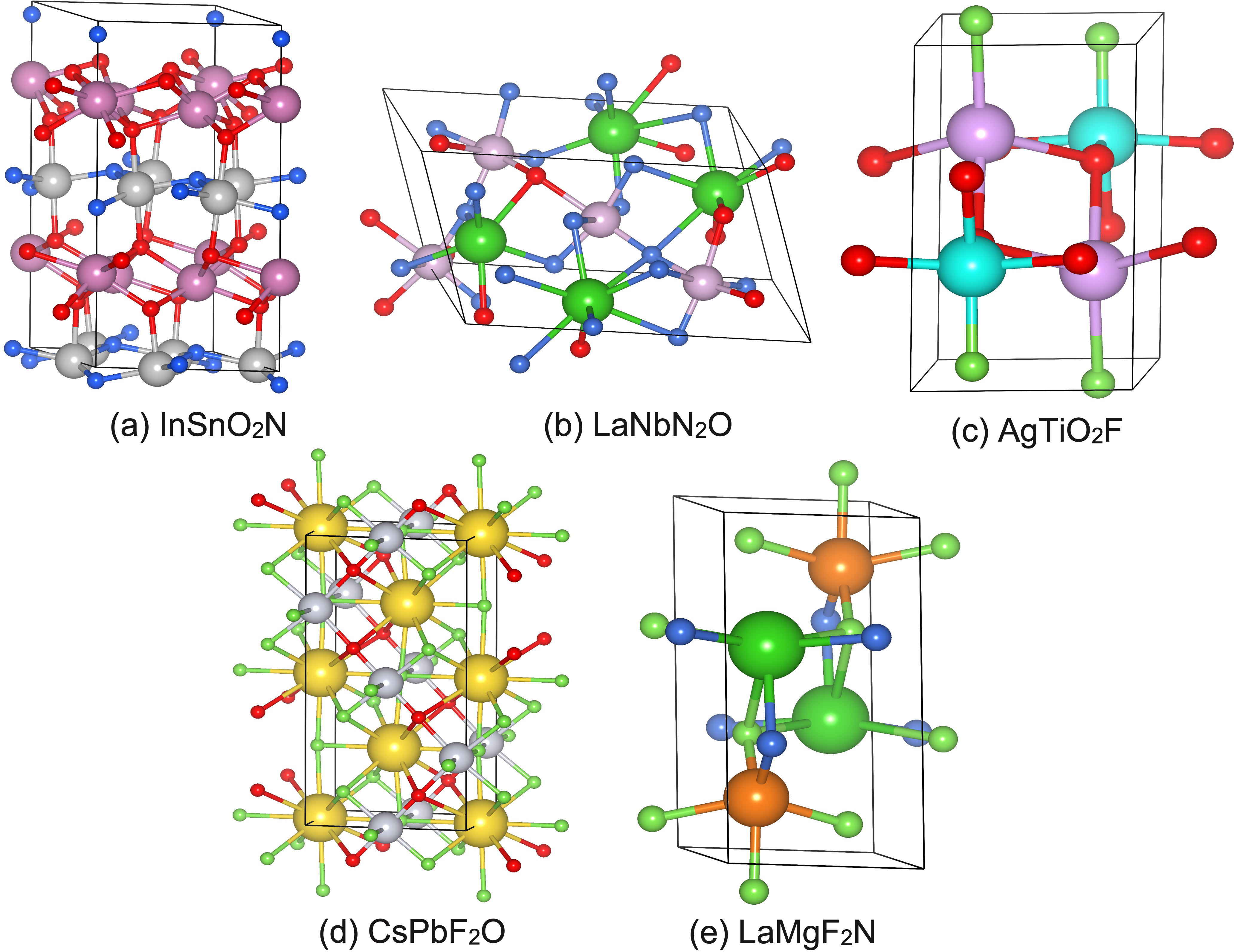}
	\caption{Optimized crystal structure of (a) InSnO$_2$N, (b) LaNbN$_2$O, (c) AgTiO$_2$F, (d) CsPbF$_2$O, and (e) LaMgF$_2$N. Images produced with VESTA~\cite{Momma:db5098}.}
	\label{1}
\end{figure}

Here, we focus on the oxynitride, oxyfluoride, and nitrofluoride perovskites i.e., ABX$_2$Y (A = Ba, Ca, La, Sr, Ag, K, Cs, Pr, In ; B = Nb, Ta, Zr, Ti, In, Pb, Mg, Sn ; and X, Y = O, N, F)
perovskites. We have considered in total 18 perovskite structures, out of which 9 are ABO$_2$N, 3 are ABN$_2$O, 4 are ABO$_2$F, 1 is ABF$_2$O and 1 is ABF$_2$N type of perovskite. The oxynitride perovskite of type ABO$_2$N e.g., InSnO$_2$N, crystallizes in a hexagonal cell and belongs to the polar space group P6$_3$\textit{cm}, as shown in Fig.~\ref{1}(a). The optimized lattice constants are a = b = 6.16 \AA\: and c = 12.26 \AA, which are well in agreement with the previous findings~\cite{doi:10.1021/acscatal.1c03737}. The perovskite type ABN$_2$O e.g., LaNbN$_2$O, crystallizes in an orthorhombic cell with lattice constants a = 5.78 \AA, b = 5.76 \AA\: and c = 10.02 \AA\: (see Fig.~\ref{1}(b)). AgTiO$_2$F oxyfluoride perovskite has tetragonal cell edges, a = b = 5.48 \AA\: and c = 7.45 \AA\: and belongs to the space group I4/mcm as shown in Fig.~\ref{1}(c). The system CsPbF$_2$O is an interesting perovskite because of the fact that the perovskite CsPbI$_3$ is the parent inorganic compound with an orthorhombic crystal structure. This perovskite has lattice constants a = 13.03 \AA, b = 4.79 \AA\: and c= 6.68 \AA. Finally, LaMgF$_2$N is the only nitrofluoride system we predict to have possibilities to be realized experimentally with space group P12$_1$/m1. The lattice constants of this perovskite structure are a = 6.33 \AA, b = 3.69 \AA\: and c = 8.25 \AA. Similarly, the lattice parameters of other perovskite structures are mentioned in SI (see section II of SI).
\subsection{Structural Stability}
\begin{figure}[h]
	\centering
	\includegraphics[width=0.45\textwidth]{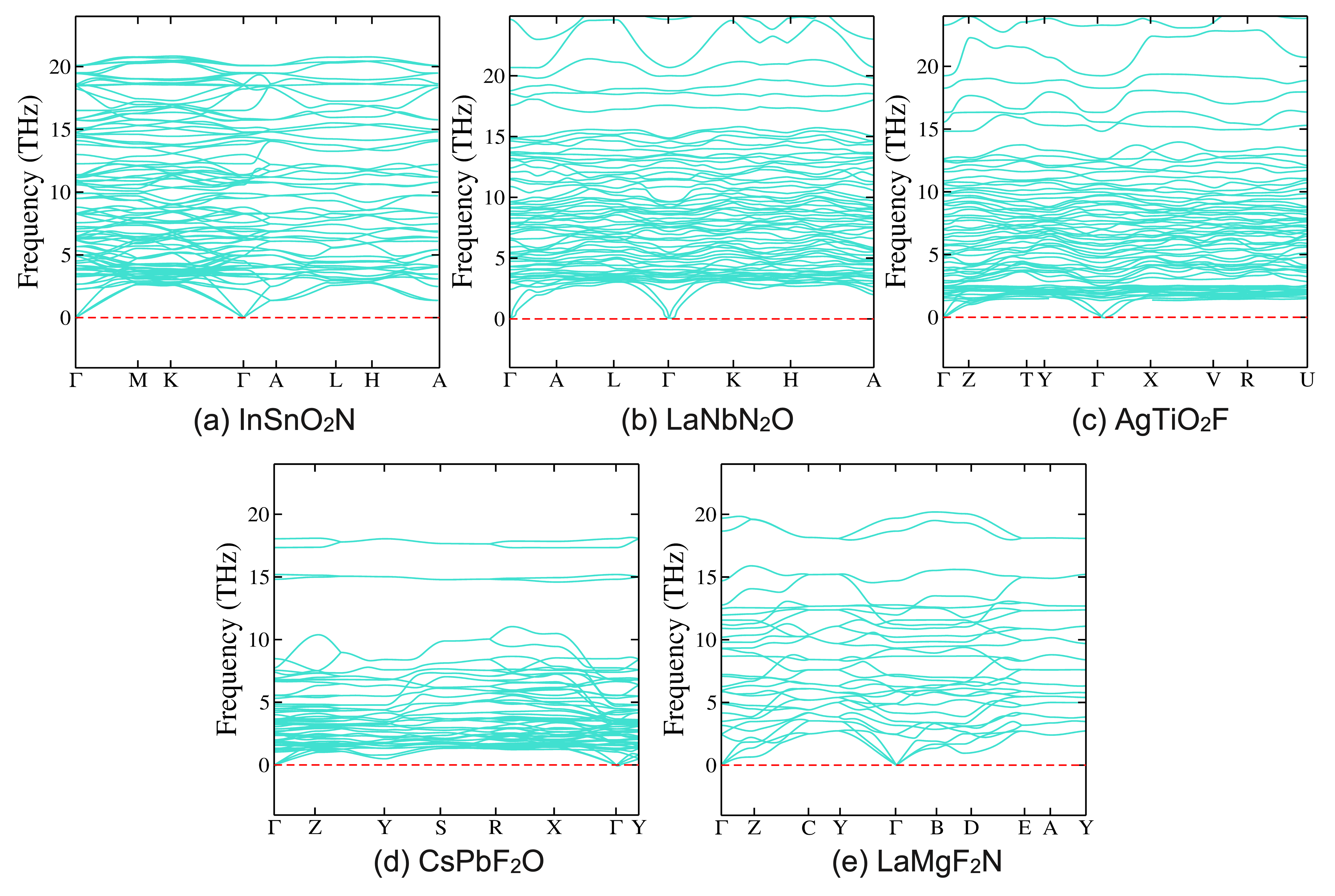}
	\caption{Phonon band structures of (a) InSnO$_2$N, (b) LaNbN$_2$O, (c) AgTiO$_2$F, (d) CsPbF$_2$O, and (e) LaMgF$_2$N.}
	\label{2}
\end{figure}

The stability of perovskites is often discussed based on the Goldschmidt tolerance factor~\cite{goldschmidt1926gesetze}
\begin{equation}
	t=\frac{r_\textrm{A}+r_{\textrm{anion}}}{\sqrt{2}(r_\textrm{B}+r_{\textrm{anion}})}
\end{equation}
where \textit{r}$_\textrm{A}$, \textit{r}$_\textrm{B}$ and \textit{r}$_\textrm{anion}$ are the ionic radii of the A and B cations, and of the anion, respectively. A more recent study based on the novel analytics approach has led to the proposition of a new factor~\cite{doi:10.1126/sciadv.aav0693}
\begin{equation}
	\tau=\frac{r_\textrm{anion}}{r_\textrm{B}}n_\textrm{A}\left(n_\textrm{A}-\frac{r_\textrm{A}/r_\textrm{B}}{\textrm{log}(r_\textrm{A}/r_\textrm{B})}\right)
\end{equation}
where \textit{n}$_\textrm{A}$ is the oxidation state of A cation. It has been reported that for a large experimental dataset of perovskites, 0.825 $<$ \textit{t} $<$ 1.08 gives a classification accuracy of 74$\%$, while $\tau$ $<$ 4.18 has an accuracy of 92$\%$. Now, to use these formulae for mixed anions, we need to decide the value of \textit{r}$_{\textrm{anion}}$ to be used. In line with the suggestions provided, we decided to use the arithmetic average of the radii of two mixed anions, i.e., \textit{r}$_{\textrm{anion}}$ = (2$\textit{r}$$_\textrm{X}$+$\textit{r}$$_\textrm{Y}$)/{3}. However, it has been noted that using \textit{r}$_{\textrm{anion}}$, the Goldschmidt tolerance factor fails to capture the stability trend of mixed anion perovskites~\cite{Pilania2020,D0DT01518K}. For this purpose, the geometric mean has also been used to approximate the radius of the mixed anions~\cite{C3TA10216E}. Also, more complicated factors like the octahedral factor and atomic packing fraction have been proposed to understand the stability of these perovskites~\cite{C3TA10216E,doi:10.1021/jacs.7b09379}. But in our case, for simplicity, we decided to consider traditional \textit{t} and $\tau$. 
\begin{table}[htbp]
	\caption{Stability parameters of ABX$_2$Y perovskites.} 
	\begin{center}
		\begin{tabular}[c]{|c|c|c|} \hline
			\textbf{ABX$_2$Y}  &  \textbf{\textit{t}} & \textbf{$\tau$} \\ \hline
			InSnO$_2$N & 0.84 & 3.32  \\ \hline
			LaNbN$_2$O  & 0.84 & 3.74  \\ \hline
			LaTaN$_2$O  & 0.84 & 3.74  \\ \hline
			PrTaN$_2$O & 0.87 & 2.56   \\ \hline
			AgTiO$_2$F  & 0.90 & 3.81  \\ \hline
			BaInO$_2$F & 0.89 & 4.20  \\ \hline
			CsPbO$_2$F & 1.07 & 3.49 \\ \hline
		    KTiO$_2$F & 0.98 & 3.52 \\ \hline
		    LaMgF$_2$N & 0.81 & 5.64 \\ \hline
		\end{tabular}
		\label{Table1}
	\end{center}
\end{table}\\
From Table~\ref{Table1}, we find that the value of \textit{t} for all the selected perovskites lies between 1.07 and 0.81, which is perfectly consistent with the usual range reported for perovskites~\cite{doi:10.1126/sciadv.aav0693}. Now concerning $\tau$, we find that all selected perovskites have value $\tau$ $<$ 4.18 except LaMgF$_2$N. For other ABO$_2$N perovskites also, we have calculated the structural stability parameters mentioned in section III of SI. In addition, to further analyze the dynamic stability, we have plotted the phonon band structures for the optimized perovskite structures. The phonon band structures calculated along the high-symmetry points of the Brillouin zone for InSnO$_2$N, LaNbN$_2$O, AgTiO$_2$F, CsPbF$_2$O, and LaMgF$_2$N are shown in Fig.~\ref{2}. Notably, the absence of negative frequencies confirms the dynamic stability of these perovskites. Similarly, we have analyzed the phonon band structures of other ABX$_2$Y perovskites in section IV of SI. 
 \\
Note that we have also checked the structural stability of different ABX$_2$Y perovskites at a higher temperature using \textit{ab initio} molecular dynamics (AIMD). We have obtained the radial distribution function \textit{g(r)} at \textit{T} = 0 K and \textit{T} = 300 K by a 6 ps long MD simulation run with \textit{NVT} ensemble (Nose-Hoover thermostat)~\cite{doi:10.1063/1.449071}. We observe that the nature of the radial distribution function for the nearest neighbors remains the same at room temperature (see section V of SI). This confirms the stability of these perovskites at operational temperature of 300 K.
\subsection{Electronic Structure}
\begin{figure}[h]
	\centering
	\includegraphics[width=0.45\textwidth]{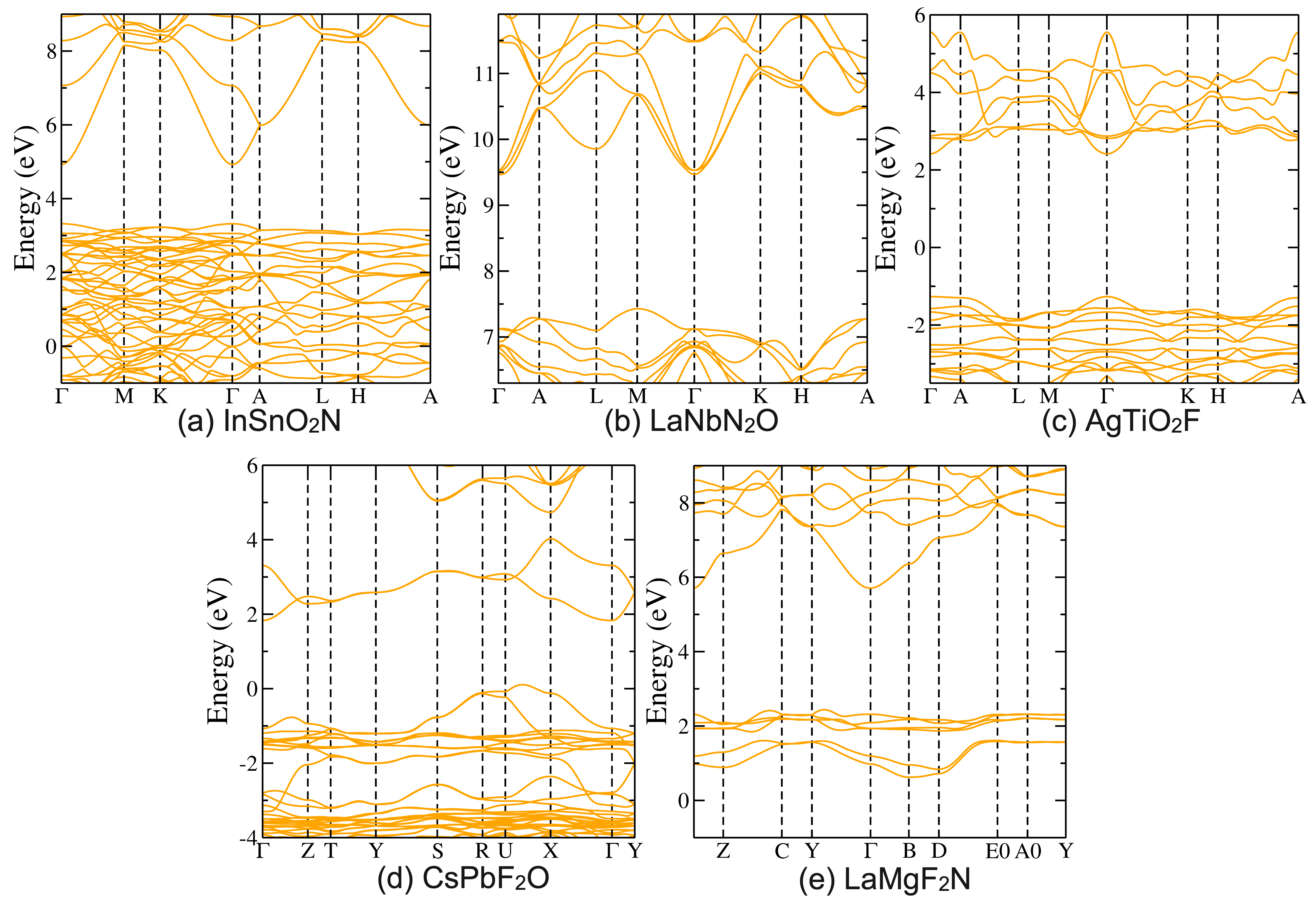}
	\caption{Electronic band structures of (a) InSnO$_2$N, (b) LaNbN$_2$O, (c) AgTiO$_2$F, (d) CsPbF$_2$O, and (e) LaMgF$_2$N using HSE06 $\epsilon_{\textrm{xc}}$ functional.} 
	\label{3}
\end{figure}

\begin{figure}[h]
	\centering
	\includegraphics[width=0.45\textwidth]{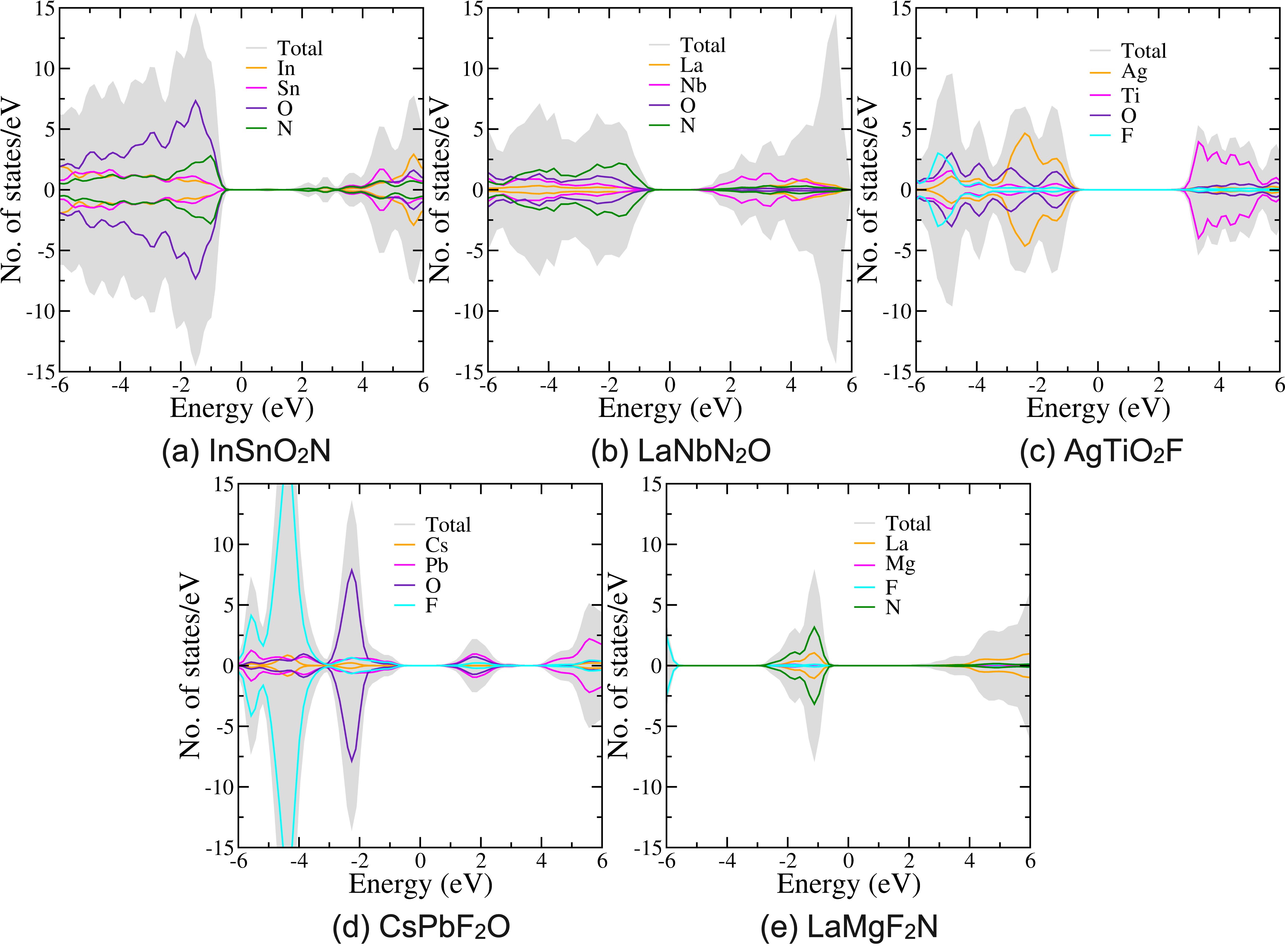}
	\caption{Partial density of states (pDOS) of (a) InSnO$_2$N, (b) LaNbN$_2$O, (c) AgTiO$_2$F, (d) CsPbF$_2$O, and (e) LaMgF$_2$N using HSE06 $\epsilon_{\textrm{xc}}$ functional.}
	\label{4}
\end{figure}

We have plotted the electronic band structure and partial density of states (pDOS) for ABX$_2$Y type of perovskites to better understand the role of A/B cations and X/Y anions near the valence band maximum (VBM) and the conduction band minimum (CBm). Fig.~\ref{3}(a) shows the band structure of InSnO$_2$N with a direct band gap of 1.60 eV at $\Gamma$ using HSE06 $\epsilon_{\textrm{xc}}$ functional. From Fig.~\ref{4}(a), we can clearly observe that the valence states are mainly composed of N and O with a small contribution coming from Sn. While the conduction states have mostly N contribution. Finally, we see very few states associated with In in the [-6, 6] eV energy window, which is compatible with the interpretation that the A atom is fully ionized in the perovskite structure.\\
The band structure of LaNbN$_2$O is rather different from that of InSnO$_2$N. It has an indirect band gap of 2.04 eV with the bottom of the CB at $\Gamma$ and the top of the VB at M (see Fig.~\ref{3}(b)). The VB is mainly contributed by N and O states, while the CB comprises of Nb states. Also, the La states are found only in the VB, indicating that this atom is ionized in this structure (see Fig.~\ref{4}(b)). For the other cases of oxynitride perovskites, all the band structures and PDOS plots are provided in section VI of SI.\\ 
From the AgTiO$_2$F band structure (see Fig.~\ref{3}(c)), we see that it has a direct band gap of 2.32 eV. Also, from Fig.~\ref{4}(c), we can clearly observe that the VB states are mostly composed of Ag and O states, while Ti states mostly contribute to the CB. The HSE06 band structure of CsPbF$_2$O is depicted in Fig.~\ref{3}(d). The band gap is 1.87 eV which is indirect, with the top of the valence and bottom of the conduction band (at $\Gamma$) composed of hybridized F, O, and Pb states. In this case, the separation of the VB into two manifolds is incomplete, leading to some overlap between the two sets of bands. The bottom of the CB, on the other hand, is separated by more than 1 eV from the rest of the CB (see Fig.~\ref{4}(d)). Similarly, for the other oxyfluoride perovskites, we have plotted the band structure and PDOS (see section VI of SI).\\
In case of nitrofluoride perovskite LaMgF$_2$N, we have obtained a large band gap of 3.31 eV. The highly dispersive bottom of the CB is mainly constructed from La states, while top of the VB has mostly N states with smaller La contribution (see Fig.~\ref{3}(e) and ~\ref{4}(e)). In LaMgF$_2$N, we also see a clear splitting of the VB. However, due to a strong electronegativity difference of N and F, the gap between the two manifolds is considerably larger than for the oxyfluoride systems.
\subsection{Optoelectronics and Dielectrics}
\begin{figure}[h]
	\centering
	\includegraphics[width=0.35\textwidth]{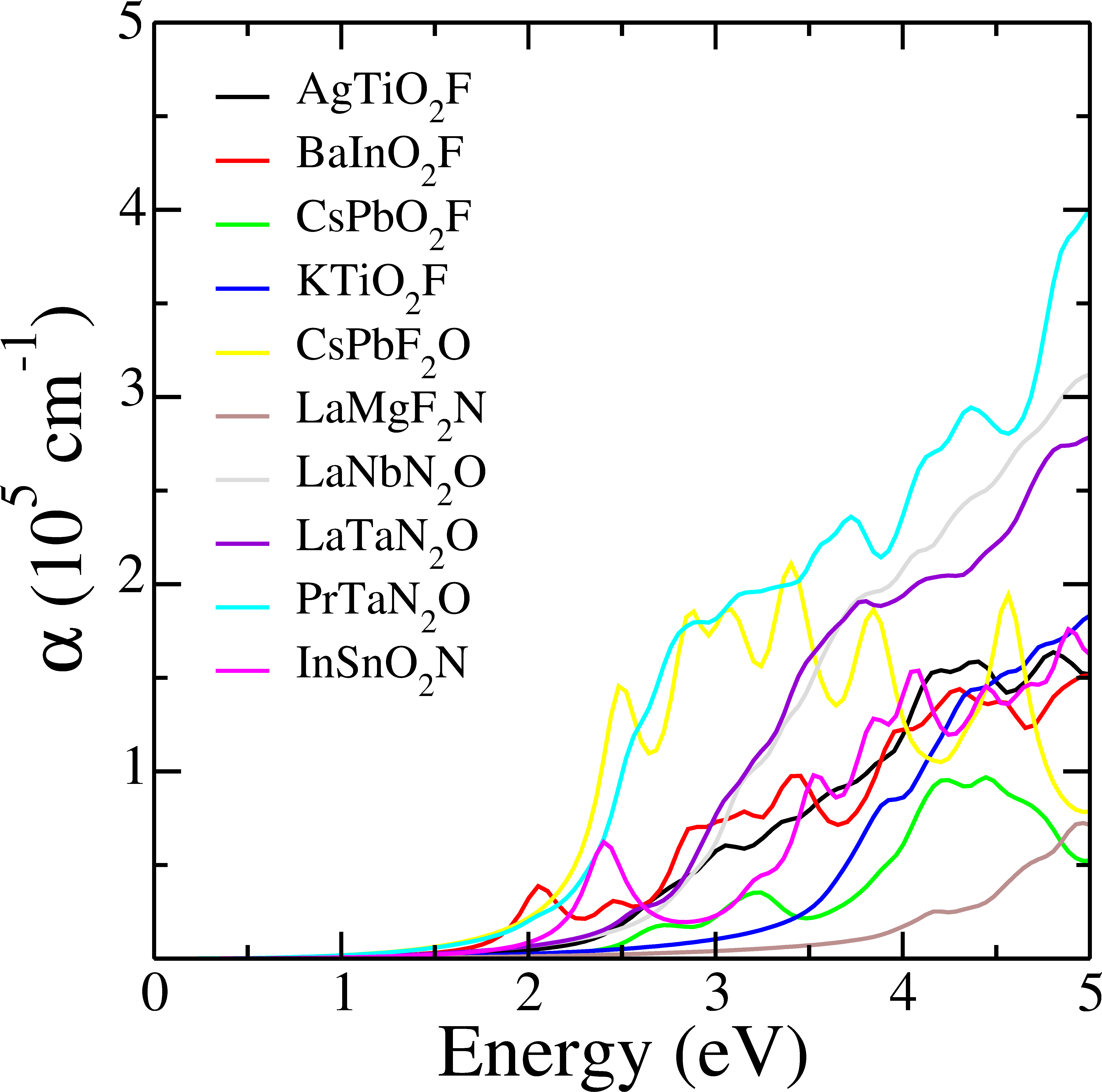}
	\caption{Absorption coefficient of ABX$_2$Y perovskites.}
	\label{5}
\end{figure}

In order to get an in-depth insight into the suitability of a particular material in optoelectronic applications, a detailed study of its optical properties like dielectric function, refractive index, extinction coefficient and absorption coefficient is indispensable. The absorption coefficient ($\alpha$) of a material is computed from the frequency dependent dielectric constant using the following formulae~\cite{D0TC01484B}, where $\varepsilon$$_1$ and $\varepsilon$$_2$ are the real and imaginary terms of the dielectric constant, respectively:
\begin{equation}
	\alpha=\frac{4\pi\kappa(\omega)}{\lambda}
\end{equation}
where $\kappa(\omega)$ is the extinction coefficient, which is given by:
\begin{equation}
	\kappa(\omega)=\sqrt{\frac{\sqrt{\varepsilon_2^2+\varepsilon_1^2}-\varepsilon_1}{2}}
\end{equation}
The absorption coefficient is one of the most essential properties of a material in terms of its photovoltaic application as it depicts key information regarding optimal solar energy conversion efficiency. The typical $\alpha$ for direct semiconductors is of the order 10$^5$~\cite{book,doi:10.1063/5.0044146}. The theoretical results show that all the perovskites exhibit high absorption coefficients (see Fig.~\ref{5} ). \\
Further, photovoltaic applications require high charge carrier mobility to reduce nonradiative recombination losses and efficiently transport the photogenerated carriers to the electrodes. In order to estimate the carrier mobility, we have calculated the effective mass of electrons and holes in ABX$_2$Y perovskites from their band structure obtained by using HSE06 $\epsilon_{\textrm{xc}}$ functional (see Fig.~\ref{3} and section VI of SI). All the perovskites have light electrons, with InSnO$_2$N having \textit{m}$_e$= 0.15\textit{m}$_0$. However, the holes are heavier, as is typical for most metal oxides, with InSnO$_2$N having a hole effective mass of \textit{m}$_h$= 1.75\textit{m}$_0$, which is in well agreement with the previous findings~\cite{doi:10.1021/acs.chemmater.0c02439}. Similarly, BaInO$_2$F has \textit{m}$_e$= 0.41\textit{m}$_0$ and \textit{m}$_h$= 0.97\textit{m}$_0$. In this case also, the hole effective mass is higher than the electron effective mass. Here, the low effective mass of the electrons in ABX$_2$Y perovskites can likely make them promising semiconductors. Also, we have calculated the electronic and ionic dielectric constant of these perovskites and found that they have a large dielectric constant value (see Table~\ref{Table2}), which are in well agreement with the previous findings~\cite{doi:10.1021/acs.chemmater.0c02439,C8TC05749D}. To this point, the relevant properties for solar energy conversion for some of the ABX$_2$Y perovskites are compiled and summarized in Table~\ref{Table2} and Table S3 of SI.\\
\begin{table}[H]
	\caption{DFT calculated properties of ABX$_2$Y perovskites. Band gap (\textit{E}$_g$), electronic dielectric constant ($\varepsilon_\infty$), ionic dielectric constant ($\varepsilon_r$), effective mass of electron (\textit{m}$_\textrm{e}^*$) and hole (\textit{m}$_\textrm{h}^*$) in terms of the rest mass of electron (\textit{m}$_0$).} 
	\begin{center}
		\begin{tabular}[c]{|c|c|c|c|c|c|c|} \hline
			\textbf{ABX$_2$Y } &  \textbf{\textit{E}$_g$ (eV)} & \textbf{$\varepsilon_\infty$} & \textbf{$\varepsilon_r$} & \textbf{\textit{m}$_\textrm{e}^*$} & \textbf{\textit{m}$_\textrm{h}^*$} \\ \hline 
			InSnO$_2$N & 1.60   & 4.45 & 10.12   & 0.15   & 1.75    \\ \hline
			LaNbN$_2$O & 2.04   & 7.27 & 61.06   & 0.33   & 0.41   \\ \hline
			LaTaN$_2$O & 2.11   & 9.28 & 70.58   & 0.36   & 0.51     \\ \hline
			AgTiO$_2$F & 2.32   & 2.74 & 39.95   & 0.82   & 1.68    \\ \hline
			BaInO$_2$F & 2.01   & 4.14 & 15.05   & 0.41   & 0.97   \\ \hline
			CsPbF$_2$O & 1.87   & 4.66 & 14.97   & 0.15   & 0.17    \\ \hline
		\end{tabular}
		\label{Table2}
	\end{center}
\end{table}
\subsection{Photocatalytic Water Splitting}
In a photocatalytic process, the perovskite material absorbs light as a result an electron at the VB is transferred to the CB, and a photogenerated e-h pair is produced. Subsequently, for the oxidation or reduction reaction, the photogenerated electrons and holes transfer to the corresponding reaction sites. The photocatalytic reaction can be described by two half-reactions for the hydrogen evolution and water oxidation which are summarized as follows:
\begin{equation}
	2\textrm{H}^+ + 2\textrm{e}^- \rightarrow \textrm{H}_2
\end{equation}
\begin{equation}
	2\textrm{H}_2\textrm{O} + 4\textrm{h}^+ \rightarrow \textrm{O}_2+ 4\textrm{H}^+
\end{equation}
For the water splitting reaction, the CBm should be higher than the reduction potential of H$^+$/H$_2$ (\textit{E}$_{\textrm{H}^+/\textrm{H}_2}$ = --4.44 eV at pH = 0.0). On the other hand, the VBM should be lower than the oxidation potential of O$_2$/H$_2$O (\textit{E}$_{\textrm{O}_2/\textrm{H}_{2}\textrm{O}}$ = --5.67 eV at pH = 0.0). From the above discussion, we conclude that in order to evaluate the feasibility of photocatalytic water splitting, it is crucial to calculate precisely the edge positions of the materials. The calculation of the band edges using DFT is quite complicated and computationally expensive because the surface chemistry and interfacial effects are considered in the process. Galli \textit{et al.} have summarized recent progress and open theoretical challenges present in simulations~\cite{Pham2017ModellingHI}. They use an appropriate method to treat the effect of interface on the band edge position of the surface structure. However, it is still complex to predict the absolute energy positions of CBm and VBM for the bulk structure.\\
Here, we have used a reliable empirical formula proposed by Xu \textit{et al.}~\cite{XuSchoonen+2000+543+556} to calculate the absolute band edges of ABX$_2$Y bulk perovskites using the following equation:
\begin{equation}
	E_{\textrm{VBM}}=\chi-\frac{1}{2}E_\textrm{g}
\end{equation}
\begin{equation}
	E_{\textrm{CBm}}=E_{\textrm{VBM}}+E_\textrm{g}
\end{equation}
where \textit{E}$_{\textrm{VBM}}$ and \textit{E}$_{\textrm{CBm}}$ represents the absolute potentials of VBM and CBm. $\chi$ is the electronegativity of the perovskite which can be determined by the absolute electronegativities of the constituent atoms as:
\begin{equation}
	\chi(\textrm{compound})=\chi_1^a\chi_2^b\chi_3^c....\chi_n^m
\end{equation}
where $\chi_1$, $\chi_2$, $\chi_3$ and $\chi_n$ represent the electronegativities of the constituent atoms; a, b, c and m are the molar fractions of the atoms. The electronegativity of the constituent atoms is calculated using the Mulliken electronegativity~\cite{doi:10.1063/1.1749394}: 
\begin{equation}
	\chi_1=(I+A)/2
\end{equation}
where I and A is the ionization energy and the electron affinity of the atom, respectively. \\
\begin{figure}[h]
	\centering
	\includegraphics[width=0.45\textwidth]{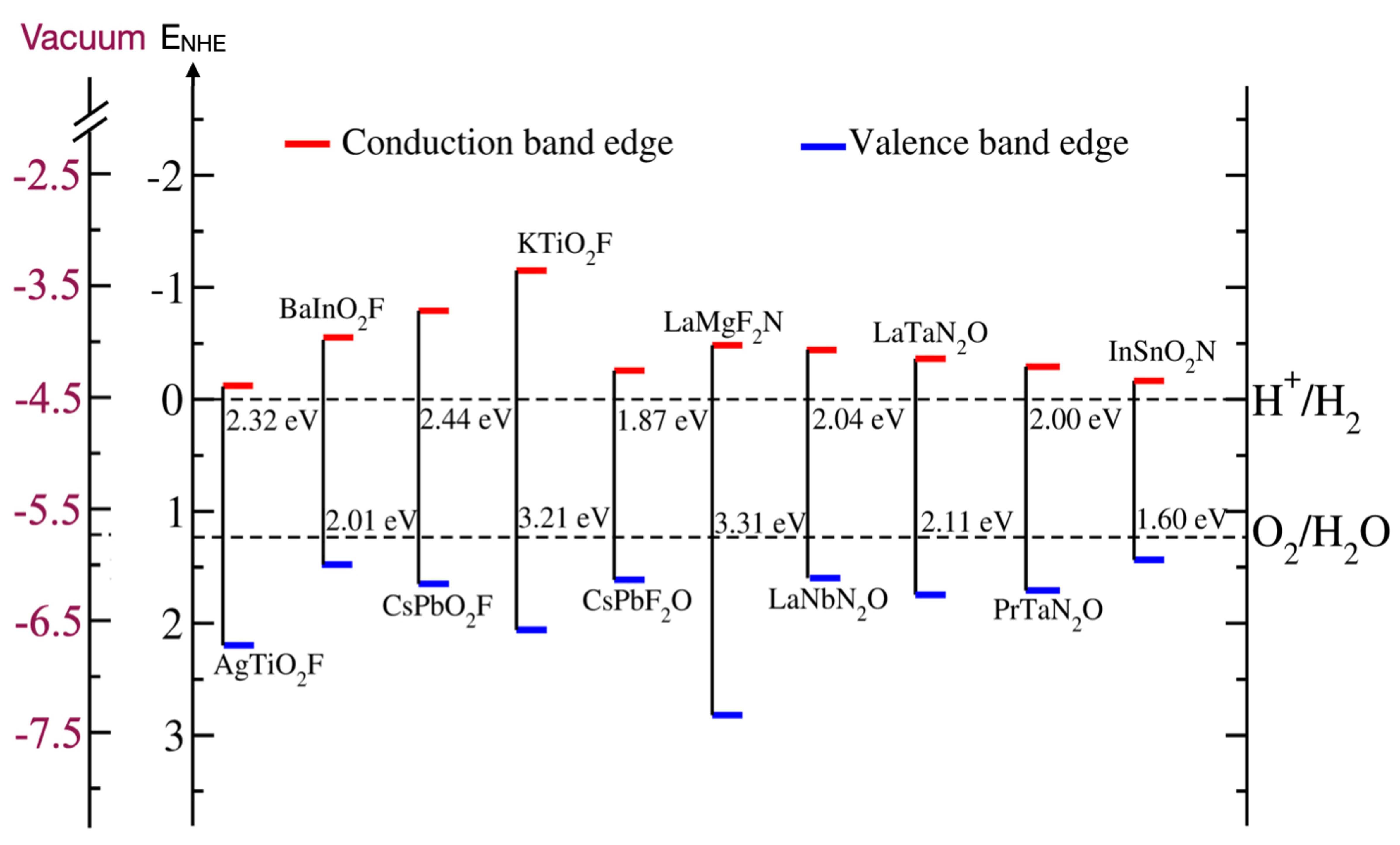}
	\caption{Band edge alignment of ABX$_2$Y perovskites w.r.t. water redox potential levels (H$^+$/H$_2$, O$_2$/H$_2$O)}
	\label{6}
\end{figure}
Fig.~\ref{6} visually expresses the absolute energy positions of band edges and the redox potentials of the water splitting reaction. In case of AgTiO$_2$F, the CBm is shifted downward by a large amount, and hence, their reduction power is very low and could not be utilized for hydrogen generation from water. For KTiO$_2$F and LaMgF$_2$N cases, the band gap is very large thus not inducing the visible-light response. In view of this, from the applicability in photocatalytic water splitting, BaInO$_2$F, CsPbO$_2$F, LaNbN$_2$O, LaTaN$_2$O and InSnO$_2$N are the most desirable ones. For other oxynitride perovskites also, we have plotted the band edge alignment (see section VIII of SI). Now to better classify these perovskites for different applications, we have discussed two applications in detail.
\subsubsection{One-photon water splitting}
\begin{figure}[h]
	\centering
	\includegraphics[width=0.45\textwidth]{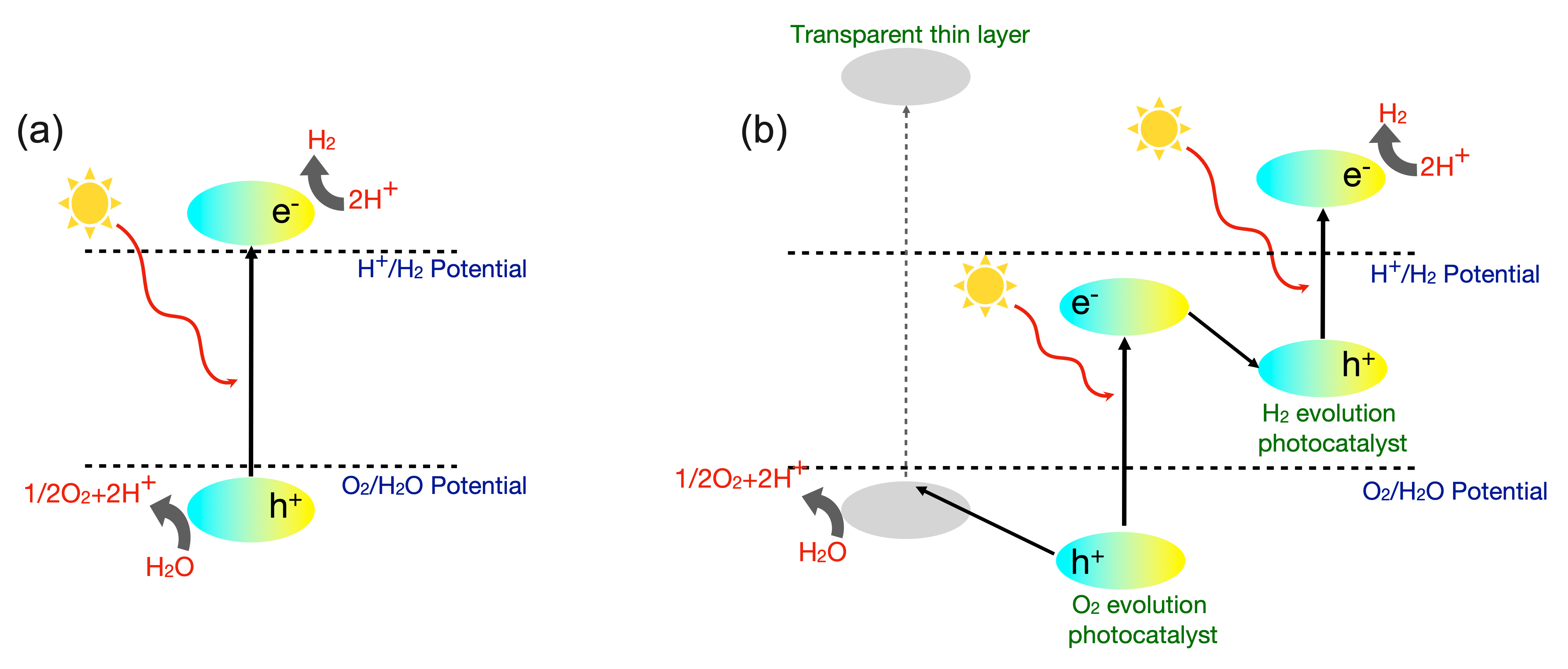}
	\caption{(a) Overall water splitting scheme, (b) Scheme of a tandem cell with a transparent protecting shield on the oxygen evolution photocatalyst.}
	\label{7}
\end{figure}

The one-photon or overall water splitting process is schematized in Fig.~\ref{7}(a). In this process, a single photon creates an e-h pair. Further, the electron and the hole reach two different regions of the surface to avoid recombination, and evolve hydrogen and oxidize water, respectively. The criteria for a material to be used for solar light capture that we have considered here are: (i) structural stability, (ii) band gap in the visible light range, i.e., 1.5 $\leq$ \textit{E}$_\textrm{g}$ $\leq$ 3 eV, and (iii) band edges straddling with the oxygen and hydrogen evolution potentials i.e., VB $>$ 1.23 eV and CB $<$ 0 eV. From Fig.~\ref{6}, we have found various ABX$_2$Y perovskites (e.g., BaInO$_2$F, CsPbO$_2$F, LaNbN$_2$O, InSnO$_2$N, etc.) which are suitable for collecting visible light and for the evolution of both hydrogen and oxygen.  
\subsubsection{Transparent shield}
Photocorrosion is one of the major problems related to the use of materials for oxygen production. In order to overcome this issue, it is required to develop a highly stable and transparent thin film to cover the oxygen evolution photocatalyst as a protective shield (see Fig.~\ref{7}(b)). For this purpose, the material should have a large band gap of, say, more than 3 eV. Fig.~\ref{7}(b) illustrates a device as a photoanode protecting transparent shield for a tandem cell. Here, the electron with higher energy evolves hydrogen, while the hole at lower energy moves through the  transparent film and evolves oxygen. The criteria for such material are: (i) a direct band gap in the UV range in order to avoid absorbing a part of the visible light spectrum, (ii) VB needs to be placed between the edges of the photocatalyst and the potential of oxygen evolution so that the hole reaches the reaction region with a small energy loss i.e., 1.7 $\leq$ {VB}$_\textrm{edge}$ $\leq$ 2.8 eV. The same idea can be used to develop a photocathode protecting shield~\cite{https://doi.org/10.1002/anie.201203585}. The only change is the position of the CB i.e., -0.7 $\leq$ {CB}$_\textrm{edge}$ $\leq$ 0 eV. From Fig.~\ref{6}, we find that KTiO$_2$F and LaMgF$_2$N can be used as a protective transparent shield.
\subsection{HER and OER mechanism}
\begin{figure}[h]
	\centering
	\includegraphics[width=0.45\textwidth]{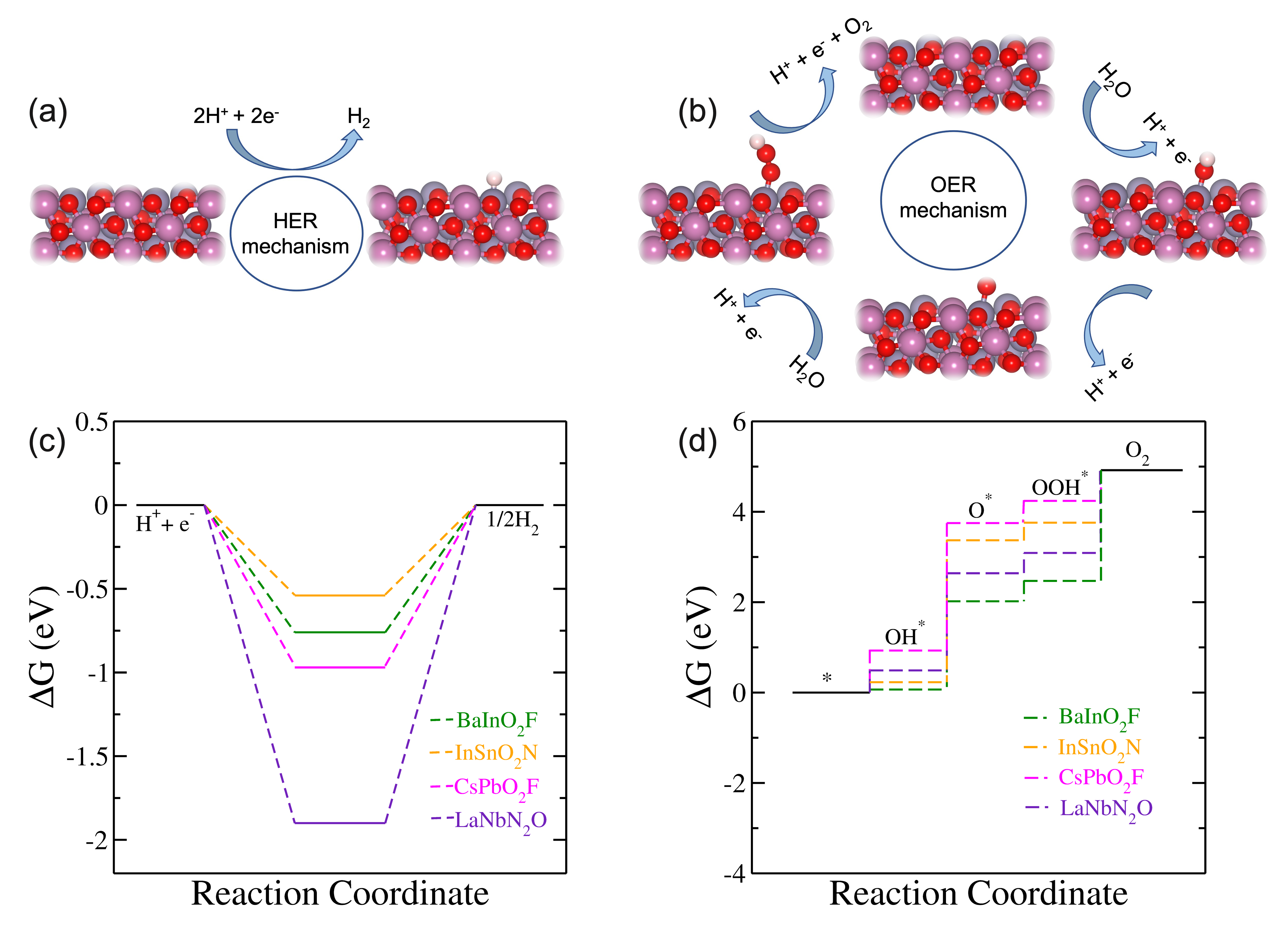}
	\caption{(a) HER steps, (b) OER steps, (c) HER free energy diagram, and (d) OER free energy diagram for ABX$_2$Y perovskites at an electrode potential E$_\textrm{SHE}$=0 V and pH=0.}
	\label{8}
\end{figure}

The HER reaction mechanism is schematized in Fig.~\ref{8}(a). In HER mechanism, the reaction steps follow: (i) adsorption of H, (ii) hydrogen reduction and (iii) desorption as H$_2$~\cite{doi:10.1021/acs.chemrev.7b00689}. The Gibbs free energy of hydrogen reaction ($\Delta$\textit{G}$_\textrm{H}$) is a reasonable descriptor for HER reaction. In order to calculate the Gibbs free energy changes ($\Delta$\textit{G}) of the intermediates in HER, the following expression is used:
\begin{equation}
	\Delta G_\textrm{H}=\Delta E_\textrm{H}+\Delta E_{\textrm{ZPE}}-T\Delta S_\textrm{H}
\end{equation} 
where,
\begin{equation}
	\Delta E_\textrm{H}=E_{\textrm{total}}-E_{\textrm{surface}}-1/2E_{\textrm{H}_2}
\end{equation} 
In the above mentioned equations, $\Delta$\textit{E}$_\textrm{H}$ is the hydrogen binding energy on the perovskite surface, $\textit{E}_{\textrm{total}}$ is the total energy for the adsorption state, $\textit{E}_{\textrm{surface}}$ is the energy of the pure surface, $\textit{E}_{\textrm{H}_2}$ is the energy of hydrogen molecule, $\Delta E_{\textrm{ZPE}}$ is the change in zero-point energy and $\Delta S_\textrm{H}$ is the difference in entropy. At 298 K, $\Delta E_{\textrm{ZPE}}-T\Delta S_\textrm{H}$ = 0.25 eV is well established in literature~\cite{doi:10.1021/acscatal.6b01211}. The HER diagram along the reaction pathway $\textrm{H}^+ + \textrm{e}^- \rightarrow \textrm{H}^* \rightarrow 1/2\textrm{H}_2$ is shown in Fig.~\ref{8}(c). Under the conditions of pH = 0 and a standard hydrogen electrode (SHE) potential of 0 V, $\textrm{H}^+ + \textrm{e}^-$ is equivalent to 1/2H$_2$. From the above discussion, we found BaInO$_2$F, InSnO$_2$N, CsPbO$_2$F and LaNbN$_2$O as probable photocatalysts.
\\
Similarly, the OER mechanism involves four proton-coupled electron transfer steps on metal ion centers with OH, O, and OOH reaction intermediates, where $*$ indicates the active metal site (see Fig.~\ref{8}(b)):
\begin{equation}
	^* + \textrm{H}_2\textrm{O} \rightarrow \textrm{$^*$OH}+ \textrm{H}^+ + \textrm{e}^-
	\label{eq1}
\end{equation}
\begin{equation}
	\textrm{$^*$OH} \rightarrow \textrm{$^*$O}+ \textrm{H}^+ + \textrm{e}^-
	\label{eq2}
\end{equation}
\begin{equation}
	\textrm{$^*$O} + \textrm{H}_2\textrm{O} \rightarrow \textrm{$^*$OOH}+ \textrm{H}^+ + \textrm{e}^-
	\label{eq3}
\end{equation}
\begin{equation}
	\textrm{$^*$OOH} \rightarrow  ^* + \textrm{O}_2 + \textrm{H}^+ + \textrm{e}^-
	\label{eq4}
\end{equation}
It has already been reported that for ABO$_3$ perovskites, mostly the B site is considered as the active site for OER reaction steps. Hence, in the case of ABX$_2$Y perovskites, we have considered the B site as the active site~\cite{C5CP02834E,C7CP06539F}. Now, a computational analysis of the OER is often simplified by only considering the $^*$OH,$^*$O, and $^*$OOH reaction intermediates appearing in the electrochemical steps, as shown in equations ~\ref{eq1}-~\ref{eq4}~\cite{ROSSMEISL200783,ROSSMEISL2005178,doi:10.1021/jp711929d,doi:10.1021/jp409373c}.\\
Following this, we have estimated the catalytic activity of a specific reaction site by the largest electrochemical step in the OER free energy profile
\begin{equation}
	\Delta G^\textrm{OER}=\textrm{max}[\Delta G_i]
\end{equation}
where $\Delta G_i$ is the change in free energy of the electrochemical OER step. Then, we derive the thermodynamic overpotential as
\begin{equation}
	\eta^{\textrm{OER}} = \Delta G^\textrm{OER}/e-1.23 V
\end{equation}
where 1.23 V is the equilibrium potential~\cite{doi:10.1021/acscatal.1c03308}. Now, the reaction steps ($\Delta G_i$) are calculated from differences in the adsorption energies of the various intermediates as given by:
\begin{equation}
	\Delta
	G_{i}=G_{\textrm{total}}-G_{\textrm{surface}}-nG_{\textrm{H}_{2}\textrm{O}}-mG_{\textrm{H}_2}
\end{equation}
where n and m are the stoichiometry coefficients that preserve the number of atoms on both sides of the respective reaction. Here, we have calculated the adsorption energies under standard conditions (pH = 0, T= 298 K) and U = 0 V using the computational hydrogen electrode, as the theoretical overpotential does not depend on the pH or the potential~\cite{doi:10.1021/jp047349j,https://doi.org/10.1002/cctc.201000397}. In this case, the energy of a proton and an electron equals half the energy of a hydrogen molecule. Fig.~\ref{8}(d) shows the free energy profile for each elementary step of OER and we have found that for each case, the overall potential is determined by $\Delta G_2$ step (for details see section IX of SI). From the above results, we obtain BaInO$_2$F, InSnO$_2$N, CsPbO$_2$F and LaNbN$_2$O as probable photocatalysts for OER.
 

\section{Conclusion} 
In summary, from our extensive computational study of structural, electronic, optical and photocatalytic properties of quaternary oxynitride, oxyfluoride, and nitrofluoride perovskites, we propose them as utmost prospective candidates for the efficient absorption and conversion of solar energy into storable fuel. Initially, using different tolerance factor and phonopy calculations, we have checked the structural and dynamical stability, respectively. After that, from the band structure and pDOS calculations, we have found that the varying electronegativity factor affects the electronic properties. Further, the optoelectronic and dielectric calculations reveal that these perovskites will enable several promising applications in solar energy harvesting through efficient solar cells and visible light optoelectronics. The results show that BaInO$_2$F, CsPbO$_2$F, LaNbN$_2$O, and InSnO$_2$N  have appropriate band edges for the overall water splitting. However, KTiO$_2$F and LaMgF$_2$N can be used as a protective transparent shield. Following this, we have calculated the theoretical overpotentials for HER and OER. Low calculated overpotentials for HER and OER suggest that certain materials may merit further study as candidates for good photocatalysts. The detailed theoretical investigation presented in this work will surely help future studies to improve the stability, engineer the band gap of perovskite absorbers and good photocatalysts for water splitting.

\section*{Conflicts of interest}
There are no conflicts to declare.

\section*{Acknowledgements}
MJ acknowledges CSIR, India, for the senior research fellowship [grant no. [09/086(1344)/2018-EMR-I]]. DG acknowledges UGC, India, for the senior research fellowship [grant no. [1268/(CSIR-UGC NET JUNE 2018)]]. SM acknowledges IIT Delhi for the financial support. SB acknowledges financial support from SERB under a core research grant (grant no. CRG/2019/000647) to set up his High Performance Computing (HPC) facility “Veena” at IIT Delhi for computational resources.

\balance
\bibliography{references2}
\bibliographystyle{rsc}
\providecommand*{\mcitethebibliography}{\thebibliography}
\csname @ifundefined\endcsname{endmcitethebibliography}
{\let\endmcitethebibliography\endthebibliography}{}
	
\end{document}


\begin{center}
{\Large \bf Supplemental Information}\\ 
\end{center}
\begin{enumerate}[\bf I.]
	
	\item Band structure plot with HSE06 and HSE06+SOC $\varepsilon_{\textrm{xc}}$ functional
	\item Optimized lattice parameters of ABX$_2$Y perovskites
	\item Stability parameters (\textit{t} and $\tau$) of ABO$_2$N perovskites
	\item Phonon band structure
	\item Radial distribution plot at 0 K and 300 K using MD simulation
	\item Electronic band structure and PDOS using HSE06 $\varepsilon_{\textrm{xc}}$ functional
	\item DFT calculated properties of ABX$_2$Y perovskites
	\item Band edge alignment of ABO$_2$N perovskites
    \item Theoretical overpotential value for OER mechanism
	
\end{enumerate}
\vspace*{12pt}
\clearpage

\section{Band structure plot with HSE06 and HSE06+SOC $\varepsilon_{\textrm{xc}}$ functional}
\begin{figure}[h!]
	\centering
	\includegraphics[width=0.95\textwidth]{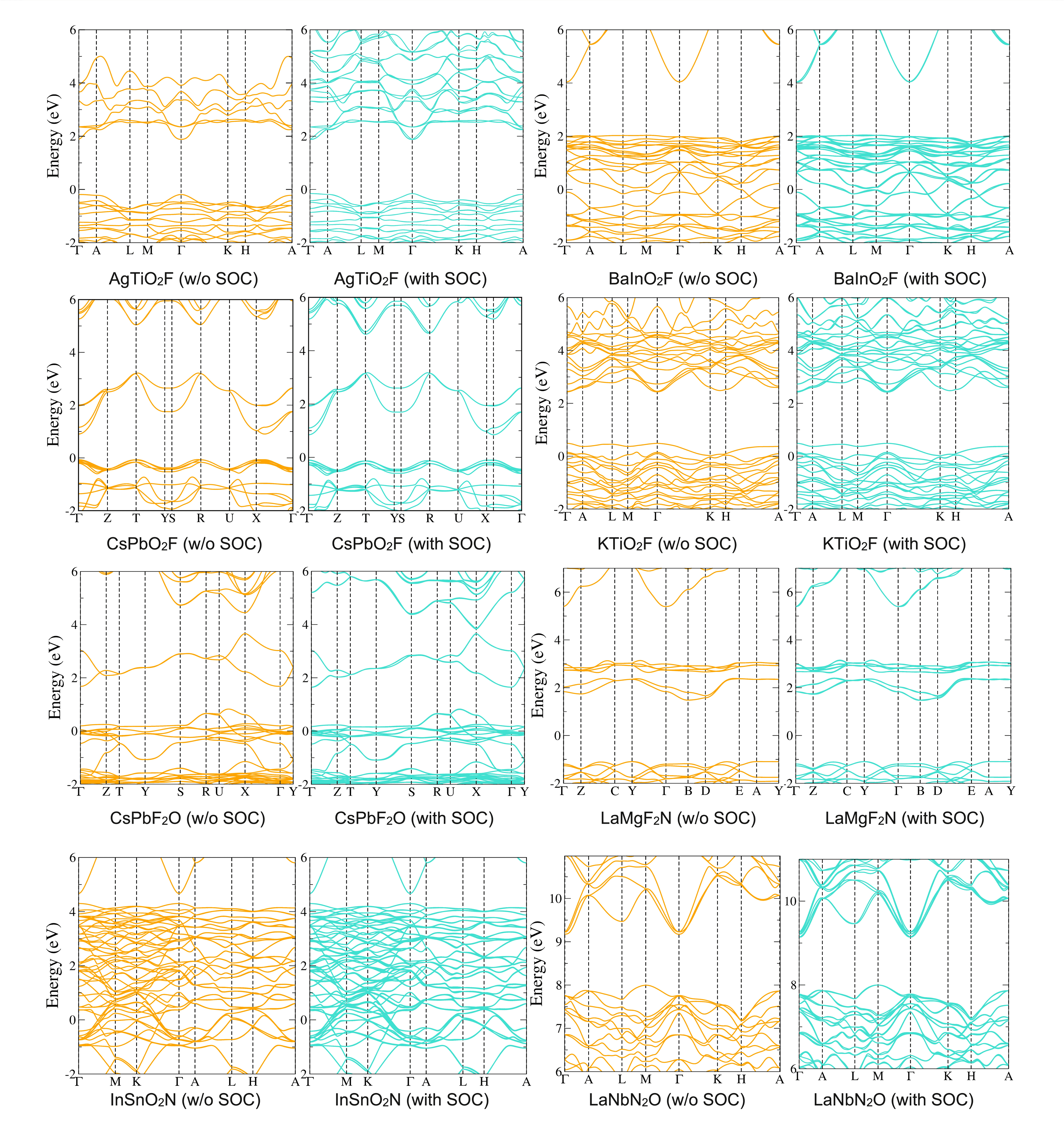}
	\caption{Band structure of ABX$_2$Y perovskites using HSE06 and HSE06+SOC $\varepsilon_{\textrm{xc}}$ functional.}
	\label{1}
\end{figure}
\newpage

\section{Optimized lattice parameters of ABX$_2$Y perovskites}

\begin{table}[htbp]
	\caption{Lattice parameters of ABX$_2$Y perovskites.} 
	\begin{center}
		\begin{tabular}[c]{|c|c|c|c|} \hline
			\textbf{ABX$_2$Y}  &  \textbf{a (\AA)} & \textbf{b (\AA)} &  \textbf{c (\AA)}\\ \hline
			BaNbO$_2$N &  5.91 & 5.96 & 10.24  \\ \hline
			BaTaO$_2$N &  5.88 & 5.88 & 7.17   \\ \hline
		    CaNbO$_2$N &  5.61 & 5.67 & 9.95   \\ \hline
			CaTaO$_2$N &  5.62 & 5.65 &	9.73   \\ \hline
	    	LaTiO$_2$N &  5.61 & 5.64 & 9.68    \\ \hline
			LaZrO$_2$N &  5.89 & 5.89 & 10.25   \\ \hline
			SrNbO$_2$N &  5.79 & 5.81 & 10.02   \\ \hline
			SrTaO$_2$N &  5.76 & 5.76 & 7.01    \\ \hline
			LaTaN$_2$O &  5.74 & 5.75 & 9.93     \\ \hline
		    BaInO$_2$F &  6.10 & 6.10 & 7.49     \\ \hline
			KTiO$_2$F  &  6.48 & 3.77 & 8.23    \\ \hline
			CsPbO$_2$F &  4.64 & 19.76 & 4.48    \\ \hline
		\end{tabular}  
		\label{Table1}
	\end{center}
\end{table}
\newpage
\section{Stability parameters (\textit{t} and $\tau$) of ABO$_2$N perovskites}
\begin{table}[htbp]
	\caption{Stability parameters of ABO$_2$N perovskites.} 
	\begin{center}
		\begin{tabular}[c]{|c|c|c|} \hline
			\textbf{ABX$_2$Y}  &  \textbf{\textit{t}} & \textbf{$\tau$} \\ \hline
			BaNbO$_2$N & 0.95 & 3.54  \\ \hline
			BaTaO$_2$N & 0.95 & 3.54  \\ \hline
			CaNbO$_2$N & 0.83 & 5.34  \\ \hline
			CaTaO$_2$N & 0.83 & 5.34   \\ \hline
			LaTiO$_2$N & 0.86 & 3.13  \\ \hline
			LaZrO$_2$N & 0.81 & 5.70  \\ \hline
			SrNbO$_2$N & 0.89 & 4.07 \\ \hline
			SrTaO$_2$N & 0.89 & 4.07 \\ \hline
		\end{tabular}
		\label{Table2}
	\end{center}
\end{table}
\newpage

\section{Phonon band structure}
\begin{figure}[h!]
	\centering
	\includegraphics[width=0.85\textwidth]{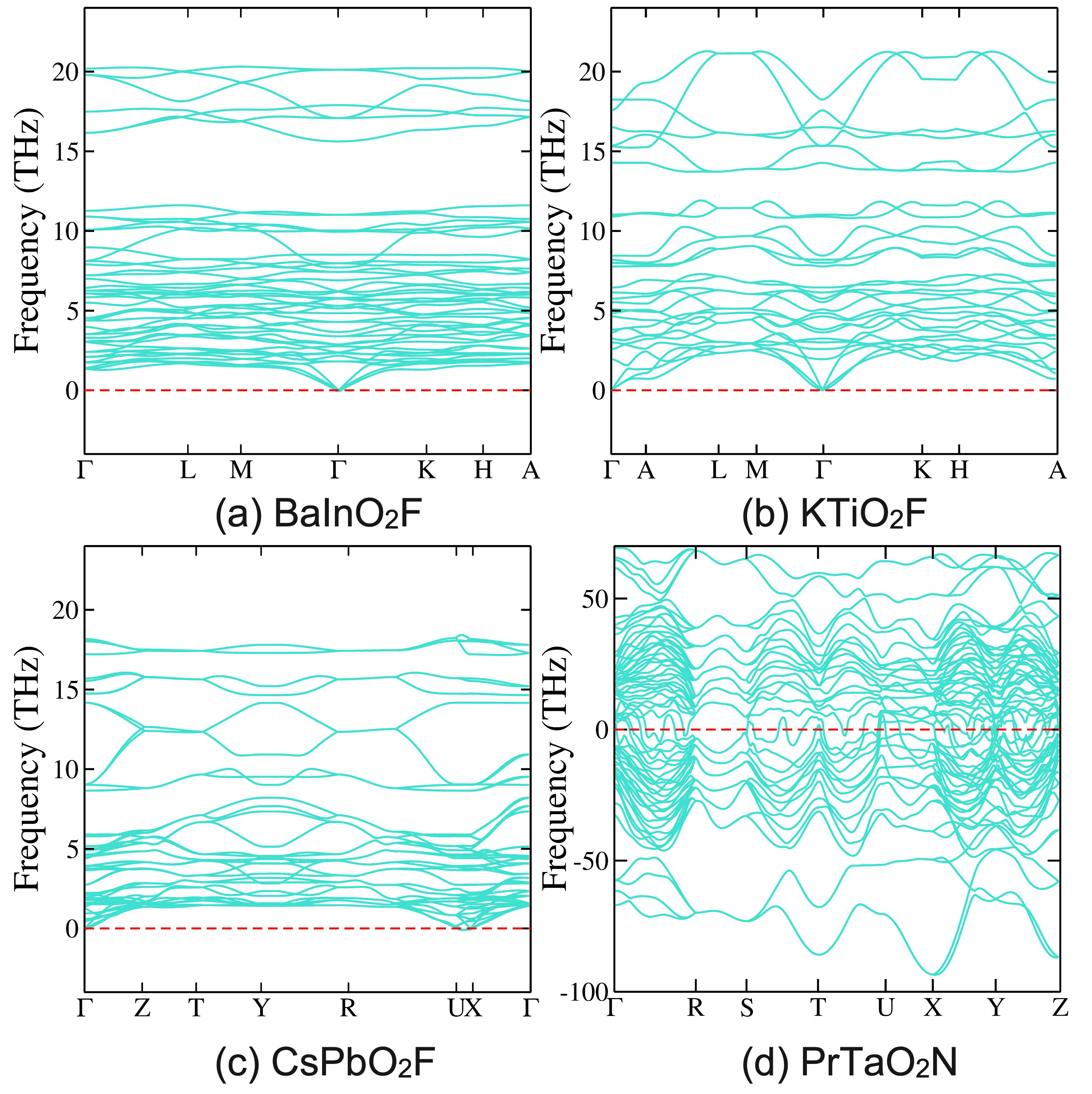}
	\caption{Phonon band structure of ABX$_2$Y perovskites.}
	\label{2}
\end{figure}
\newpage
\section{Radial distribution plot at 0 K and 300 K using MD simulation}
\begin{figure}[h!]
	\centering
	\includegraphics[width=0.56\textwidth]{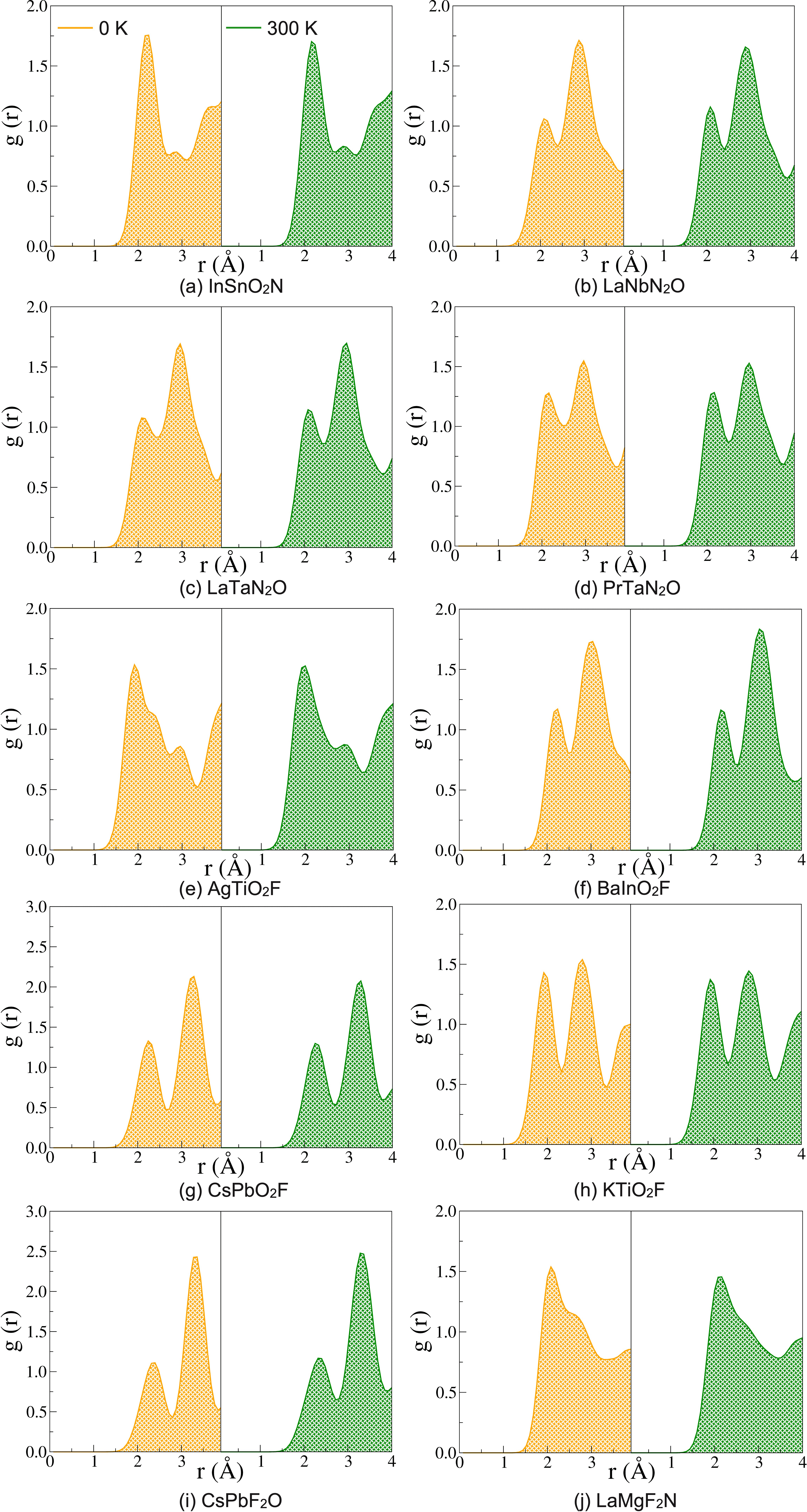}
	\caption{Radial distribution function for different ABX$_2$Y perovskites at T = 0 K and T = 300 K.}
	\label{3}
\end{figure}
\newpage
\section{Electronic band structure and PDOS using HSE06 $\varepsilon_{\textrm{xc}}$ functional}
\begin{figure}[h!]
	\centering
	\includegraphics[width=1.00\textwidth]{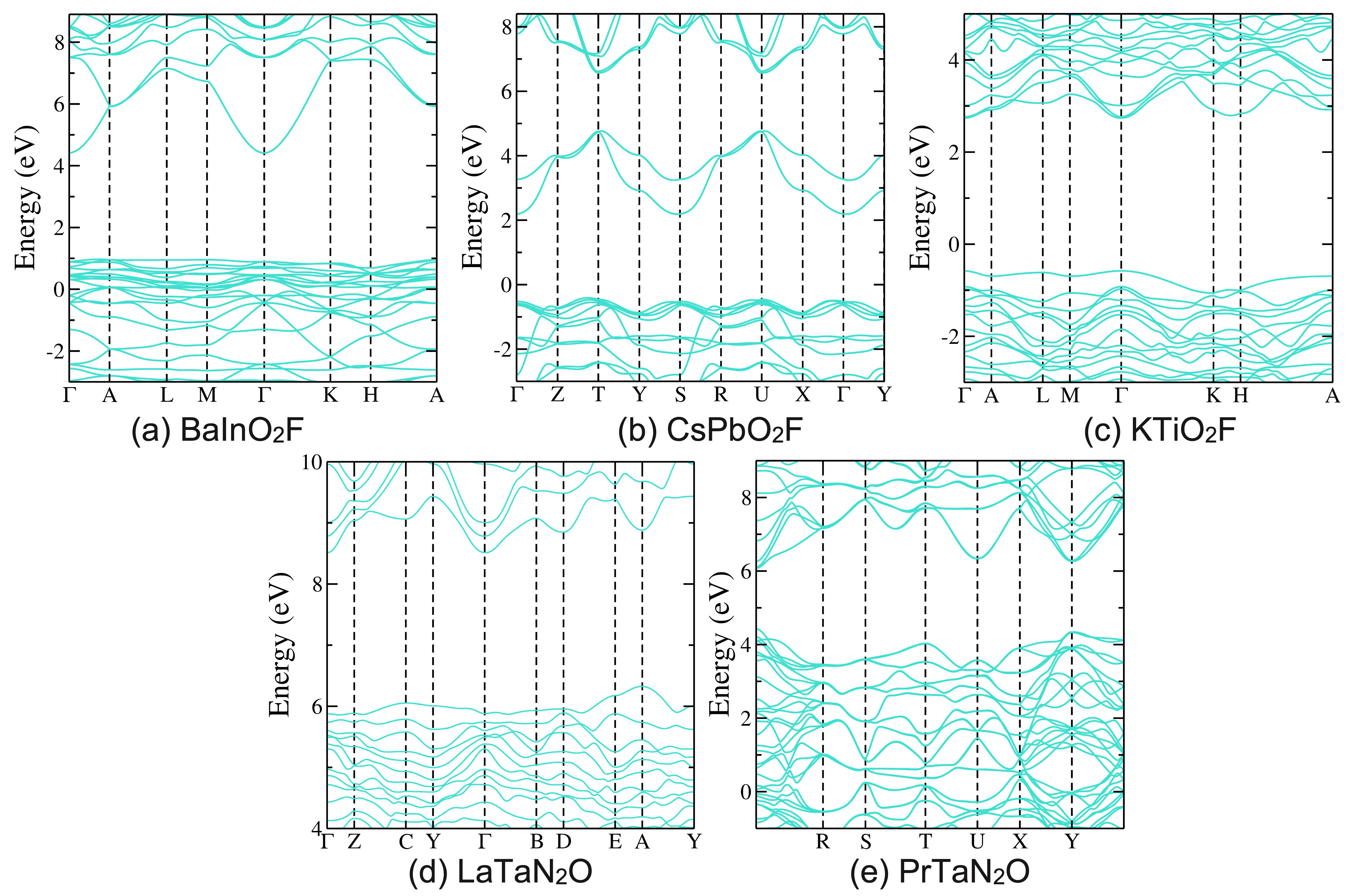}
	\caption{Band structure of (a) BaInO$_2$F, (b) CsPbO$_2$F, (c) KTiO$_2$F, (d) LaTaN$_2$O and (e) PrTaN$_2$O perovskites using HSE06 $\varepsilon_{\textrm{xc}}$ functional.}
	\label{4}
\end{figure}
\newpage

\begin{figure}[h!]
	\centering
	\includegraphics[width=1.00\textwidth]{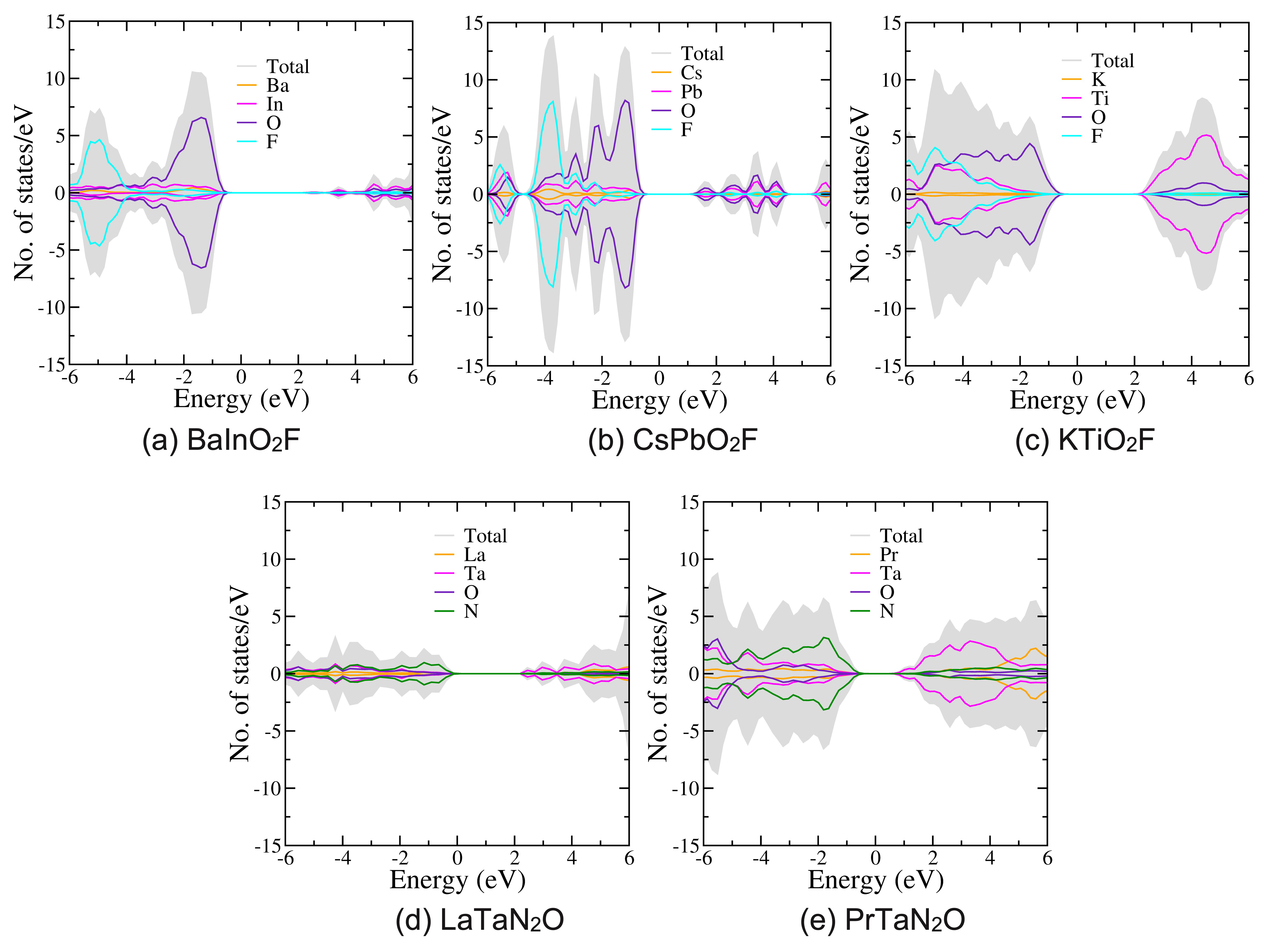}
	\caption{PDOS of (a) BaInO$_2$F, (b) CsPbO$_2$F, (c) KTiO$_2$F, (d) LaTaN$_2$O and (e) PrTaN$_2$O perovskites using HSE06 $\varepsilon_{\textrm{xc}}$ functional.}
	\label{5}
\end{figure}
\newpage

\section{DFT calculated properties of ABX$_2$Y perovskites}
\begin{table}[htbp]
	\caption{DFT calculated properties of ABX$_2$Y perovskites. Band gap (\textit{E}$_g$), electronic dielectric constant ($\varepsilon_\infty$), ionic dielectric constant ($\varepsilon_r$), effective mass of electron (\textit{m}$_\textrm{e}^*$) and hole (\textit{m}$_\textrm{h}^*$) in terms of the rest mass of electron (\textit{m}$_0$).} 
	\begin{center}
			\begin{tabular}[c]{|c|c|c|c|c|c|c|} \hline
				\textbf{ABX$_2$Y } &  \textbf{\textit{E}$_g$ (eV)} & \textbf{$\varepsilon_\infty$} & \textbf{$\varepsilon_r$} & \textbf{\textit{m}$_\textrm{e}^*$} & \textbf{\textit{m}$_\textrm{h}^*$}  \\ \hline 
				BaNbO$_2$N & 1.70   & 6.88 & 34.59   & 0.55   & 0.39     \\ \hline
				BaTaO$_2$N & 1.46   & 7.07 & 54.89   & 0.79   & 1.27    \\ \hline
				CaNbO$_2$N & 1.94   & 6.68 & 27.25   & 0.70   & 1.21    \\ \hline
				CaTaO$_2$N & 2.00   & 6.42 & 32.31   & 0.49   & 0.87     \\ \hline
				LaTiO$_2$N & 1.64   & 8.75 & 62.56   & 0.72   & 1.00   \\ \hline
				LaZrO$_2$N & 2.65   & 6.47 & 56.44   & 0.63   & 0.77     \\ \hline
				SrNbO$_2$N & 1.66   & 7.08 & 40.86   & 0.61   & 2.24   \\ \hline
				SrTaO$_2$N & 1.72   & 6.77 & 43.46   & 0.99   & 0.60   \\ \hline
			\end{tabular}
		\label{Table3}
	\end{center}
\end{table}
\newpage

\section{Band edge alignment of ABO$_2$N perovskites}
\begin{figure}[h!]
	\centering
	\includegraphics[width=1.00\textwidth]{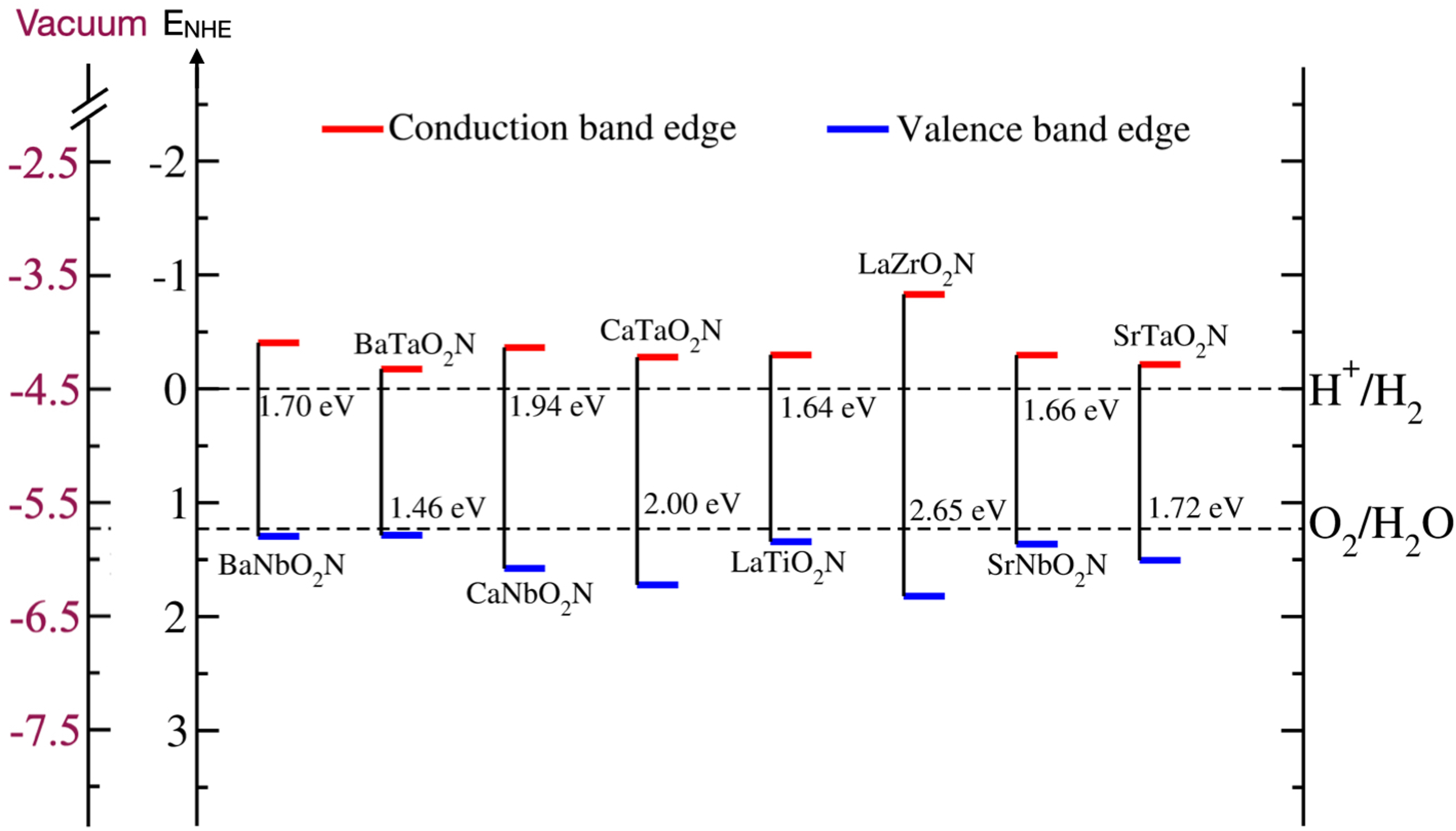}
	\caption{Band edge alignment of ABO$_2$N perovskites w.r.t. water redox potential levels (H$^+$/H$_2$, O$_2$/H$_2$O).}
	\label{6}
\end{figure}
\newpage

\section{Theoretical overpotential value for OER mechanism}
\begin{table}[htbp]
	\caption{Calculated value of overpotential of ABX$_2$Y perovskites.} 
	\begin{center}
		\begin{tabular}[c]{|c|c|} \hline
			\textbf{ABX$_2$Y } &  \textbf{$\eta^{\textrm{OER}}$ (V)} \\ \hline 
			InSnO$_2$N & 1.91      \\ \hline
			LaNbN$_2$O & 0.92    \\ \hline
			BaInO$_2$F & 0.71    \\ \hline
			CsPbO$_2$F & 1.58       \\ \hline
		\end{tabular}
		\label{Table4}
	\end{center}
\end{table}
\newpage